\documentclass[11pt]{article}

\usepackage{graphicx} 

\begin{document}

\noindent\LARGE{\textbf{Adsorption of small aromatic molecules on gold: a DFT localized basis set 
study including van der Waals effects} }
\vspace{0.6cm}

\noindent\large{\textbf{Luiza Buimaga-Iarinca \textit{$^{a}$} and
Cristian Morari$^{\ast}$\textit{$^{a}$}}}\vspace{0.5cm}

\vspace{0.6cm}

\noindent \normalsize{
We compare the density functional theory (DFT) results on the adsorption of small aromatic molecules (benzene, pyridine and thiophene) on gold surfaces
obtained by using three types of van der Waals exchange-correlation functionals and localized basis set calculations.
We show that the value of the molecule-surface binding energy depends on the interplay between the BSSE effect and 
the tendency of the exchange-correlation
functionals to overestimate both the molecule-surface as well as the gold-gold distances within the relaxed systems.
Consequently, we find that by using different types of LCAO basis sets or geometric models for the adsorption of 
the molecules on the surface,
the binding energy can vary up to 100 \%. A critical analysis of the physical parameters resulting from the 
calculations is presented for each exchange-correlation functional.}
\vspace{0.5cm}

\section{Introduction}

\footnotetext{\textit{$^{a}$~National Institute for Research and Development of Isotopic and Molecular Technologies, 
Department of Molecular and Biomolecular Physics, 65-103 Donath, 400293 Cluj-Napoca, Romania, 
cristian.morari@itim-cj.ro}}

The physical and chemical properties of molecules at surfaces play an important role in various fields, ranging from heterogeneous catalysis to nanoscale mechanics and organic optoelectronics \cite{Waser}. 
While the properties of individual molecules can be modified in a controlled manner by chemical synthesis, their binding at surfaces can give rise to new functionalities, which may be used to design surfaces with specific properties \cite{Balzani, Cuniberti, k10, Morari, Luiza, RSC}. 

Simple aromatic molecules such as benzene, thiophene, or pyridine can serve as molecular building blocks for larger molecules used in organic
electronics \cite{Ma}. Consequently, the adsorption of these molecules on several noble metal surfaces has been studied in detail both experimentally and 
computationally \cite{gross, Grimme, Bilic1, Xi, Zhou, benzen_exp, Dudde, Su, Matsuura, Lipkowski, Milligan, Liu1, Nambu, Demuth}.
Experiments have shown that benzene interacts weakly with surfaces of noble metals \cite{Xi, Zhou, benzen_exp}. On (111) surfaces it adsorbs in a flat-lying
manner and forms ordered 3 $\times$ 3 overlayers \cite{Dudde}. Heteroaromatic molecules, such as thiophene or pyridine, were found to undergo a potential or coverage driven phase transition when they are adsorbed on noble metal surfaces from solution \cite{Su, Matsuura, Lipkowski} or under high vacuum conditions \cite{Milligan, Liu1, Nambu, Demuth}. In the low coverage regime the heteroaromate adopts 
a flat-lying conformation, while upon increasing the coverage a tilted or vertical orientation was observed. 

In spite of these intensive investigations, it was difficult to reproduce correctly 
the binding energies or the flat-lying orientations of small aromatic molecules at density functional theory (DFT) level \cite{gross} until recently. One of the major reasons for these difficulties is the correct description of the long-range van der Waals interactions in the frame of DFT.

Even though van der Waals forces are weak compared to ionic or covalent bonds they can play a fundamental role in nanotechnology applications \cite{nature1, nature2}. This is mostly caused by the fact that van der Waals forces are anisotropic (they depend on the relative orientation of the interacting molecules and/or surfaces). They also proved to be responsible for a large fraction of the molecule-surface interactions; for example, according to Aradhya et al. \cite{nature2},
the contribution of the dispersion forces in the process of breaking a molecular junction is surprisingly large. More precisely they found that dispersion forces are comparable or even larger than the forces needed to break an $Au-N$ bond \cite{nature1, nature2}.

Quantitative estimations of the importance of van der Waals effects in the DFT studies were recently pointed out in literature \cite{Mura, felice, Medeiros, Wellendorf, Yldirim}. 
DFT is a rich source of information for the understanding of the complex interplay between the forces arising at the molecule-surface interface \cite{Morari,Xue,Storhoff,Park,Harnack,Parak}. 
By means of the DFT models, it is possible to perform a detailed description of the geometric structure of the small molecules adsorbed on gold surfaces.
Nevertheless, the DFT is limited by the quality of the employed exchange-correlation functional.
Standard functionals (such as PBE \cite{PBE}) tend to underestimate the adsorption energies, if compared to the experimental data \cite{DNA_vw2,DNA_vw3}.
It was shown recently that the computed binding energies and geometric properties obtained by taking into account long-range interactions indicate significant improvement of the results, suggesting that their use in the study of various adsorbed molecules is of outmost importance \cite{Klimes, Mura, felice, gross,Medeiros,Wellendorf,Yldirim,Chwee,Nadler,Abbin,Tonigold}. 

Among the recent results on the role played by van der Waals interaction in adsorption of aromatic systems on various surfaces let us remind the study of Medeiros et al \cite{Medeiros} which emphasized that the binding energy per carbon atom in benzene, coronene and acicurmocoronene adsorbed on gold follows a linear trend.
They pointed out that this trend can reproduce the binding energy of graphene on gold. 
The studies of Yildirim and Kara \cite{Yldirim} for the adsorption of olympicene on Cu(111) revealed the unavoidable importance of the self-consistent van der Waals interactions for an accurate description of adsorption characteristics of organic species on metal substrates, also providing a minimal list of characteristics to be considered for distinguishing the weak physisorption from the strong chemisorption \cite{Yldirim}. Self-consistent inclusion of long range interactions via opt-type functionals \cite{KBM} was found to change significantly the adsorption heights, energies and charge transfers, compared with those obtained from standards GGA functionals \cite{Yldirim}. 

Several studies on the adsorption of small aromatic molecules exist \cite{gross,Toyoda,liu,Wellendorf}, based on various approaches such as DFT-D \cite{gross, Grimme}, QM-MM hybrid \cite{gross}, non-self consistent DFT \cite{Toyoda, Wellendorf}.
Rigorous studies performed by using plane waves also showed very good agreement with experimental data \cite{Wei1}.
While very accurate calculations were done, it is important from the practical point of view to investigate the problem by using localized basis sets (LCAO). The methods based on LCAO present the advantage of investigating with good accuracy systems with large number of atoms.
Our aim is to provide a theoretical insight on the effect of van der Waals interaction upon the adsorption mechanism of test systems, represented by three small aromatic molecules on Au(111) surface, by using the recent developments in the field of van der Waals corrected exchange-correlation functionals \cite{DRSLL, LMKLL, KBM, Perez} and LCAO basis sets \cite{Soler, Siesta}. We analyze the geometric properties, binding energies and the molecule-surface charge transfer. A comparison between the results obtained for each exchange-correlation functional
is presented for the above mentioned properties.

\section{Computational details}

DFT is employed as the first principles quantum mechanical method. It provides
efficient and typically accurate estimates of the molecule-crystal binding energies, molecular dissociation, surface atom rearrangements, and the overall electronic structure of a nanoscopic system \cite{Mura, felice, gross,Medeiros,Wellendorf,Yldirim,Chwee,Nadler,Abbin,Tonigold}.

It is well-known that DFT with a plane-wave basis set properly models electronic properties of bulk materials with 
Bloch-like wave functions spreading over the entire unit cell. For low dimension systems, such as molecules or 
surfaces the use of plane waves can became very expensive in term of computing time. Such systems can be 
more efficiently described by using localized basis sets. 

First-principles DFT calculations reported here were performed using the SIESTA 
method \cite{Soler,Siesta} that combines the advantages of localized basis sets with those of periodic systems analysis. 
Among the disadvantages we can recall the BSSE effect \cite{BB}. Specifically, since SIESTA uses localized basis 
sets, the overlap of the gold basis set 
with the orbitals used as basis set for the organic atoms, leads to an artificial decrease of the
total energy. This energy shift can be removed by using the counterpoise correction \cite{BB} in order to get results as accurate as those obtained by using plane waves for the adsorption of molecules on metallic surfaces \cite{Lee, Chapman}.
Nevertheless, it has been suggested that specific conditions have to be fulfilled in order to reach a high degree of accuracy \cite{Lee}. (The specifics for benzene adsorbed on gold will be discussed latter on.) Overall, we found that the BSSE effect can lead to a severe overestimation of the binding energy if the basis set is not properly chosen.
Therefore, we found useful to inspect the values of the binding energies without the BSSE corrections included. If these binding energies are too different from the experimental results, the BSSE corrections are expected to lead to unphysical results (even to non-bonded systems). In the following we performed detailed investigations only for selected system (see next section).

In our calculations we used three  exchange-correlation functionals dedicated to the computation of van der Waals interactions \cite{LMKLL, DRSLL, KBM}. 
The first option was the vdw-DF2 functional of Lee et. al. \cite{LMKLL} (LMKLL); next we used the vdw-DF functionals of Dion et. al. \cite{DRSLL} (DRSLL) and the functional of Klime\v{s} et. al. \cite{KBM} (KBM).

Core electrons were replaced by norm-conserving pseudopotentials in their fully non-local formulation \cite{PPS1}. 
For each type of functional a specific set of pseudopotentials was generated and tested by starting
from the parameters for the GGA-PBE functional \cite{PPS_DATA}. For gold, relativistic corrections were used
for all functionals. A uniform mesh with a plane-wave cutoff of 300 Ry was used for integrations in the 
real space \cite{Soler}. We used a standard double-zeta polarized (DZP) basis set for gold atoms atoms. 
For organic atoms we use a triple-zeta polarized (TZP) quality basis set.

Periodic boundary conditions were used to describe the infinite close packed 
face-centered cubic (fcc) gold surfaces. The spurious periodicity in the direction
perpendicular to the surface is suppressed by the large size (i.e. 32 \AA) of the cell along the Z-coordinate, imposing a vacuum layer of about 25 \AA\ between the periodic images of the systems in the Z-direction. 
The size of the cell along the X and Y axes is dictated by the requirement of isolating the molecule from its periodic replicas. Thus, the simulation cell contains 100 Au atoms, consisting of a four-layer 5 $\times$ 5 periodically repeated slab for the Au(111) surface. 
We used a 2$\times$ 2 $\times$ 1 Monkorst-Pack grid for the integrals in 
the Brillouin zone. The standard implementation of SIESTA was used to compensate 
the electric field created by the dipole moment of the system due to the asymmetry of the slab \cite{Siesta}.

Isolated molecules in the gas phase were treated employing a cubic cell of 25 $\times$ 25 $\times$ 25 \AA\ and $\Gamma$ point calculation.
The binding energy molecule-surface $\Delta E$ is defined as the difference between the total energy of the molecule-surface composite and the sum of the total energies
of the free relaxed molecule and the surface:
\begin{equation}
 \Delta E = E_{M/S} - (E_{M} + E_{S})
\label{eq:binding}
\end{equation}
where  $E_{M/S}$ is the total energy of the molecule-surface composite, $E_{M}$ is the total energy of the free relaxed molecule and $E_{S}$ is the energy of the relaxed surface.
The negative values of $\Delta E_{i}$ imply a stable adsorption composite systems \cite{BSSE}.

In order to get the final geometric structures we allow a full relaxation of the molecules adsorbed on top of the metal surface up to a  maximum gradient set to 0.01eV/\AA. 

\section{Results and discussion}

\subsection{Optimization of the computational model}

We start by computing the  relevant geometrical parameters for the clean gold surface and for the free molecules respectively.
Precisely, we compute the value of the bulk parameter for gold ($a_0$) and compare it with the experimental value (4.08 \AA\ ). 
For the isolated molecules we also compare the geometric parameters (such as bond lengths and bond angles) with the experimental ones; in addition, we compute the dipole moments 
for pyridine and thiophene, which are expected to play important role in the adsorption mechanism
(the experimental values are 2.21 D and 0.53 D respectively \cite{dipol1, dipol2}). 

\begin{table*}
\caption{Summary of the results for the tests on the three exchange correlation functionals: dipole moments $ {\cal P} $ for thiophene (${\cal P}_{T})$
and pyridine (${\cal P}_{P}$)
bulk parameter ($a_0$) and dipole moment of the relaxed gold surface ($ {\cal P}$ ).
Binding energy of benzene on Au(111) surface ($\Delta E$) and molecule-surface distance ($D$)
are given for each of the geometrical models described in the text (labeled 1 to 3). The dipole moments are given in Debye, distances are given in \AA\
and binding energies in eV.}

\label{tbl:1}
\begin{tabular}{c|rr|r|rrr|rrr}
\hline
& ${\cal P}_{T}$ &${\cal P}_{P}$ &a$_0$ & $\Delta E_1$ & $\Delta E_2$ & $\Delta E_3$ & $ D_1$ &$ D_2$ &$ D_3$  \\
\hline
\hline
LMKLL& 0.40 & 2.23 & 4.26 & -2.86 & -3.07 & -3.28 & 3.09 & 2.97 & 2.77  \\
DRSLL& 0.41 & 2.20 & 4.15 & -1.80 & -2.04 & -2.11 & 2.95 & 2.83 & 2.74 \\
KBM  & 0.40 & 2.22 & 4.18 & -2.13 & -2.48 & -2.62 & 2.90 & 2.72 & 2.50 \\
\hline
\end{tabular}
\end{table*}

In Table 1 we report the results obtained after testing different pseudo-atomic orbitals (PAO) basis sets in SIESTA \cite{Siesta}.
For gold we use a DZP basis set with an energy shift of $\delta \epsilon =350$ meV, while for organic atoms we use a TZP basis set
with $\delta \epsilon = 15$ meV. In order to get these values we performed previous tests for values of $\delta \epsilon$ (not reported here) and basis set quality (i.e. DZP or TZP). Precisely, for gold we scanned values from 300 meV to 380 meV; we found that by decreasing the value of $\delta \epsilon$ (i.e. by using long-ranged PAOs) the resulting bulk 
parameters are increasingly larger.
Since for $\delta \epsilon = 300 $ meV we found values between 4.16 \AA\ (DRSLL) up to  4.31 \AA\ for LMKLL
we did not performed tests for smaller values of $\delta \epsilon$. 
For organic atoms we test values of the energy shift between 10 meV and 150 meV.  

It can be seen from Table 1 that DRSLL leads to the best results for the bulk parameter of gold. The computed value
of 4.15 \AA\ is only slightly overestimated with respect to experimental one (0.07 \AA\ larger).  In the case of 
LMKLL the 
overestimation is the most important, leading to a difference of 0.18 \AA\ , compared to the experimental value. 
By comparing the dipole moments obtained with each of the three functionals we note that all results 
are close to the computational accuracy.  

In order to optimize our computational approach we relax the molecule-surface systems and compute the binding energy $\Delta E_{i}$  
by using different geometric models to model the molecule-metal structure, as described below.

As a first model  we use the experimental value of the bulk parameter $a_0=4.08$ \AA\ to build the gold slab. 
Geometric properties of the clean slab are determined by relaxing the position of the
top two layers of gold atoms, while the atoms in the two deep layers are pinned to their bulk
positions. The benzene molecule is adsorbed on this relaxed surface and its position is furthermore relaxed
by keeping fixed all gold atoms. 
The second model is similar to Model 1, but we used as bulk parameter for gold the values obtained from the DFT calculations
(see Table 1). Finally, the third model is similar to Model 2, but in this case we also allow 
the relaxation of the top two layers of gold in the presence of molecule \cite{model3} .

The resulting  binding energies  for the these models are summarized in Table 1 (we remind that the experimental value 
is -0.63 eV \cite{benzen_exp}). No BSSE corrections were applied to these results. 
It can be clearly seen that all values  are severely overestimated, for all exchange-correlation functionals. By applying 
a BSSE correction, we can expect an improvement, but we think that for such an important BSSE correction  
to the final results may be highly questionable. Also, it was pointed out that in the case of dispersion-dominated systems the
counterpoise correction for BSSE have the tendency to be too large \cite{Sherrill}. Therefore, we can expect  
an error cancellation, possibly leading to realistic results. Consequently, we decide not to follow this strategy
and we conclude that the use of a basis set that reproduce good bulk parameters for gold (i.e. short-range numerical orbitals) is a wrong strategy
to follow in order to describe the adsorption of aromatic molecules on gold. 
As a final remark, we note that for DRSLL we get values that are closest to the experimental results.
This corroborates with the fact that the bulk parameter for gold was estimated with the smallest error for DRSLL.

By comparing the geometric properties resulting from our calculations with experiment we note that the 
benzene-surface distance is reproduced rather well (we note that experimental values are 
around  2.9 - 3.0 \AA\ \cite{Toyoda, Abad})
Indeed, the molecule-surface distances are close to 3 \AA\ for all models. The largest value is 3.09 \AA\ obtained
by using Model 1 and LMKLL, while the smallest value for the molecule surface (i.e. 2.50 \AA\ ), is obtained in Model 3 by using KBM functional. 

We conclude that the for the first three models the molecule-surface distance reproduces 
correctly the
experimental results, while the binding energies are all wrong. The cause for the wrong binding energies is the presence of 
a very important BSSE effect.

For one last geometric model (Model 4) we choose to keep frozen the positions of gold atoms in the surface 
surface (i.e. we use the positions of the gold atoms in an ideal surface with bulk parameter $a_0 =4.08$ \AA\ ) 
and we optimize the orbitals for 
gold  (i.e. by using long-range PAO's).
We note that keeping the atoms from the surface fixed is an approach that was also previously used in similar
context \cite{Toyoda, Medeiros}.

The binding energies obtained (with and without BSSE corrections) for different values of the energy shift $\delta \epsilon $
are listed in Table 2. First, it can be seen that the values of the binding energy clearly display a convergence trend
for all functionals around  the value of $\delta \epsilon \approx 15$ meV. 
Also, the BSSE corrections diminishes for  small values of the energy shift; for $\delta \epsilon \approx 15$ their value is around 0.1 eV while
for $\delta \epsilon > 100$ meV the BSSE corrections are larger than 0.5 eV.
Second, the converged values of $\Delta E$ including BSSE corrections are between -0.3 and -0.4 eV for all three functionals. In other words, they are
all underestimated with respect to the experiment or to the plane-wave results \cite{Toyoda, Wei1} . The differences between the three functionals are small,
with the best results for LMKLL functional. 

This underestimation of binding energy is probably due to the absence of a dipole moment of the surface, that generally arise from
the relaxation of top layers in the gold slab. This is expected to lead to a slight increase of the binding energy but while in out model all gold atoms were pined up to their position, 
the surface dipole moment was not taken into account.

Also, we can speculate that the structural relaxation or the
molecule-surface charge transfer may be slightly influenced by the BSSE effect. We'll discuss this late aspect in detail
in the section dedicated to the analysis of the molecule-surface charge transfer.

To conclude, by fixing the positions of gold atoms on the one hand and using long range PAOs on the other hand we get
realistic results. Moreover, it can be seen that by using short range PAOs, unrealistic (i.e. too large) values of the BSSE
corrections are necessary which makes the final results highly questionable. 

\begin{table}
\caption{ Binding energy of benzene on Au(111) ($\Delta E$)for different LCAO basis sets,
(represented by value of the energy shift ($\delta \epsilon$) used to define the PAOs) 
for the exchange-correlation functionals. All vales are in eV. For each system
we report the values with (left) and without (right) BSSE corrections.}

\label{tbl:2}
\begin{tabular}{c|rrr}
\hline
$\delta \epsilon $ [eV] & LMKLL  &  DRSLL & KBM  \\
\hline
\hline
0.005              &  -0.40/-0.55 & -0.30/-0.41 & -0.35/-0.45  \\
0.015              &  -0.39/-0.55 & -0.29/-0.68 & -0.33/-0.49  \\
0.05               &  -0.36/-0.73 & -0.27/-0.70 & -0.28/-0.59  \\
0.10               &  -0.30/-0.99 & -0.24/-0.77 & -0.22/-0.80  \\
0.15               &  -0.22/-1.33 & -0.14/-0.88 & -0.09/-0.95  \\
0.20               &  -0.14/-1.66 &  0.01/-1.02 &  0.07/-1.14  \\
\hline

\end{tabular}
\end{table}

Finally, the average distances molecule-surface for all functionals 
are listed in Table 3. A clear trend can be seen; precisely, large values of  $\delta \epsilon $
(i.e. short-ranged PAOs) lead to smaller values for the molecule-surface distance.
For a $\delta \epsilon$ close to 5 - 15 meV the molecule surface 
reaches the value of 3.75 \AA\ . Such values are slightly larger than those reported by Wei et al. for plane wave 
calculations \cite{Wei1} (i.e. 3.24 to 3.43 \AA\ ) but are  close to the value reported by Toyoda (i.e. 3.7 \AA\ ).
Nevertheless, we note that in all cases the molecule-surface distance was overestimated.
To conclude, Model 4 gives reasonable good results for binding energies, while the geometric properties
of the molecule-metal systems are poorly described. This is the opposite situation with respect to
the first three models, where we found a good description of the geometric properties for molecule-surface 
distance but wrong values for binding energy. 

\begin{table}
\caption{Average molecule-surface distance ($D$) (in \AA\ ) for different LCAO basis sets,
(represented by value of the energy shift, $\delta \epsilon$) 
and for all investigated exchange-correlation functionals.}

\label{tbl:3}
\begin{tabular}{c|rrr}
\hline
 $\delta \epsilon $ [eV] & LMKLL  & DRSLL & KBM   \\
\hline
\hline
0.005  & 3.73 &  3.82 &  3.78       \\
0.015  & 3.72 &  3.89 &  3.77  \\
0.05   & 3.59 &  3.81 &  3.59  \\
0.10   & 3.51 &  3.68 &  3.51  \\
0.15   & 3.42 &  3.48 &  3.32  \\
0.20   & 3.34 &  3.31 &  3.19  \\
0.25   & 3.29 &  3.10 &  3.08  \\
0.30   & 3.13 &  3.01 &  3.01  \\
0.35   & 3.11 &  2.90 &  2.81  \\
\hline
\end{tabular}
\end{table}

\subsection{Geometry of the Adsorbates }

By using the computational model described in the previous section 
as Model 4 (gold atoms are frozen to bulk positions), we studied the adsorption of benzene (B), pyridine (P) and thiophene (T) molecules
on Au(111) surface. We used an energy shift $\delta \epsilon = 15$ meV for all atoms.
Previous data available in literature \cite{gross, bilic2} described the preferred adsorption configurations       
for the three molecules; precisely, for benzene and thiophene a nearly flat geometry was pointed out
while for the pyridine a slightly vertical orientation was emphasized.
It was also found \cite{gross} that lateral translation of the molecule
from the optimized structure leads to adsorption sites that are close in energy with respect to the 
minimum of the interaction energy (up to 40 meV \cite{gross}). Since such values
are close to the computational accuracy of DFT, we start our geometry optimization by a preliminary investigation
of the potential energy surface of the molecule - metal for each system. 

We start by generating geometric structures that are close
to those reported in literature \cite{gross}.
Next, we  produce a set of 25 structures by translating the molecules parallel to the surface.
Precisely, the molecules are translated along the unit cell of the gold surfaces,
by splitting the unit cells into a 5 x 5 mesh of points.
For each system we compute
the total DFT energy, in order to get a perspective of the energy landscape 
for molecule-metal interaction. 
The systems with the lowest energy where then relaxed as described in the previous two sections.

In addition to the flat orientation of the pyridine and thiophene with respect to the surface
(we symbolize these structures as -P$_\parallel$ and  T$_\parallel$ throughout the paper) 
we also investigate the structures with these molecules adsorbed at 
a perpendicular orientation with respect to the surface (i.e. P$_\perp$ and  T$_\perp$). 

The final structures are represented in figure \ref{fig:structures} together with a contour plot
of $\Delta \rho(\vec r)$ defined in equation \ref{d_rho} (see below). While in general we found positions similar to those
described in \cite{gross} we notice that for pyridine adsorbed on Au(111) we found as the most stable a position
with the nitrogen atom lying on a hollow position on the surface. 

\begin{figure}
\includegraphics[width=8cm, angle=0]{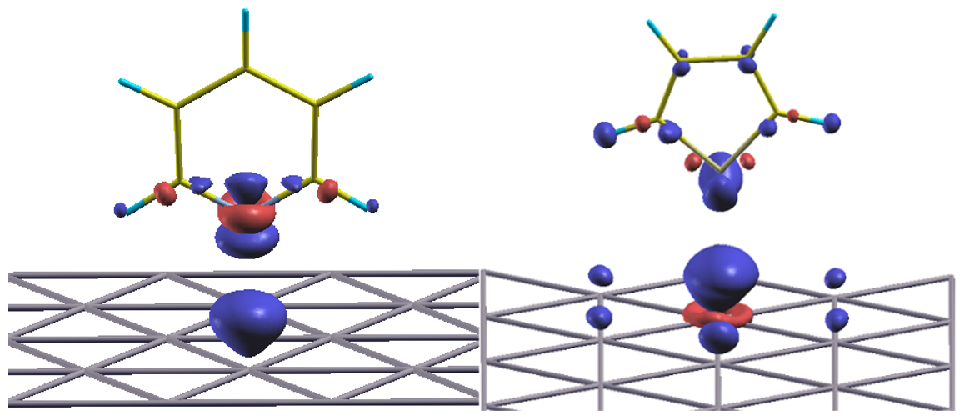}
\includegraphics[width=8cm, angle=0]{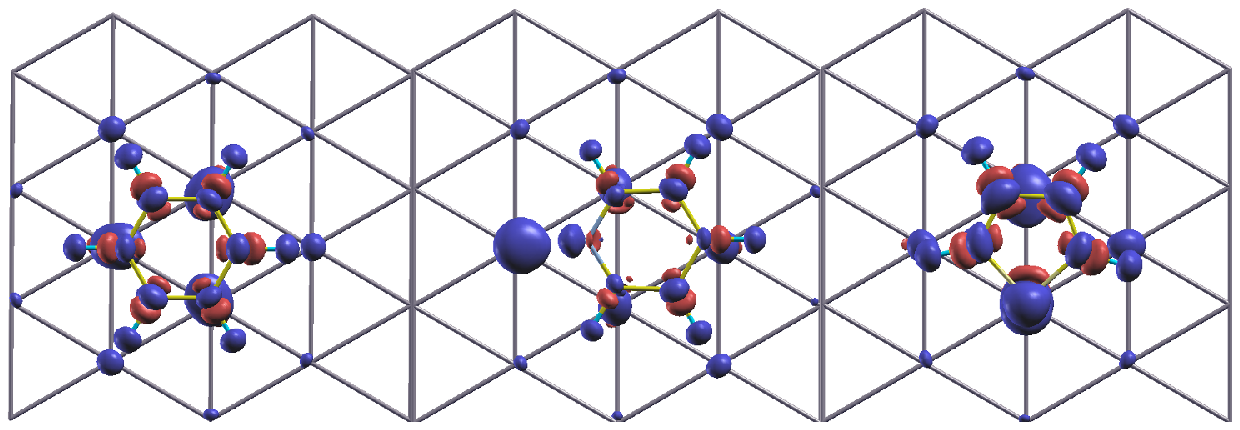}
\caption{Graphical representation of the relaxed structures together the contour plot of  $\Delta \rho (\vec r) $ at
0.0005 e/Bohr$^3$. For pyridine oriented perpendicular we used a value of 0.005 e/Bohr$^3$. 
Top: pyridine and thiophene (perpendicular orientation).
Bottom: benzene, pyridine and thiophene (flat orientation), pyridine and thiophene (perpendicular orientation).}
\label{fig:structures}
\end{figure}

Selected values describing the geometry of the molecule-surface system for the relaxed structures are summarized in Table 4.
We note that in order to label the atoms in pyridine and thiophene we number the carbon atoms clockwise, starting from the nitrogen (sulphur) atom.
By inspecting the largest versus the smallest molecule-surface distances we notice that benzene on Au(111) display a
nearly perfect parallel orientation with the surface. For flat-lying pyridine occur differences of about 0.02-0.06 \AA\
between minimum and maximum distance with respect to the surface. The most important of these differences occurs for
the KBM functional (0.06 \AA\ ) while for LMKLL the flat orientation is favored (only 0.02 \AA\ difference).
For thiophene the differences are up to  0.22 \AA\ (for DRSLL) indicating a slightly tilted orientation of the molecule with respect to the surface.

For perpendicular orientations of the last two molecules we remark that KBM favors a geometry with the molecules close to
the surface, while the opposite is valid for DRSLL.

Let us now discuss the variation of the carbon-nitrogen (sulphur)-carbon bond angles in pyridine and thiophene in the adsorbed configurations
by computing the values of $\delta \alpha = \alpha_{free} -\alpha_{adsorbed}$. 
We noticed that the values of $\delta \alpha$ for flat orientations are close to zero, as expected 
(practically no deformation of the molecule occurs in the adsorbed molecules).
For perpendicular orientation, the most important value occurs for pyridine (-1.41 degree). While this is
a non-negligible value, its influence on the energetic stability of the molecule is reduced.
We also analyzed the variation of bond lengths upon the adsorption and we found that all variations are less than 0.01 \AA\ . This
is no surprise, since little or no chemical interaction is expected to take place between molecules and surface.

\begin{table*}
\caption{Geometric parameters of the adsorbates: minimum and maximum distances from molecule's atoms 
(indicated in parentheses) to the surface ($D_{min}$ and $D_{max}$), maximum variation of a 
 bondangle ($\delta \alpha$) in the adsorbates. From top to bottom, results for LMKLL, DRSLL and KBM.}

\label{tbl:4}
\begin{tabular}{c|rrrrr}
\hline
                 & B/Au  & P$_\parallel$/Au& P$_\perp$/Au & T$_\parallel$/Au &T$_\perp$/Au \\
\hline
\hline
$D_{min}$ [\AA]  &  3.72 (C) & 3.68 (C2) & 2.80 (N)& 3.62 (C2) & 3.48 (S) \\
$D_{max}$  [\AA] &  3.73 (C) & 3.70 (N)  & 5.60 (C3)& 3.75 (S) & 6.05 (C2) \\
$ \delta \alpha $ [deg] & $-$& -0.13    & -1.22    & -0.04     & -0.37 \\
\hline
$D_{min}$ [\AA]  &  3.91 (C) & 3.93 (N) & 2.90 (N)& 3.81 (C2) & 3.47 (S) \\
$D_{max}$  [\AA] &  3.92 (C) & 3.96 (C3)  & 5.51 (C3)& 4.03 (S) & 6.02 (C2) \\
$ \delta \alpha $ [deg] & $-$ & -0.07    & -1.26    & -0.03     & -0.37 \\
\hline
$D_{min}$ [\AA]  &  3.77 (C) & 3.87 (C2) & 2.66 (N)& 3.76 (C2) & 3.37 (S) \\
$D_{max}$  [\AA] &  3.77 (C) & 3.93 (N)  & 5.43 (C3)& 3.87 (S) & 5.90 (C2) \\
$ \delta \alpha $ [deg] & $-$ & 0.0    & -1.41    & 0.04     & -0.45 \\
\hline

\end{tabular}
\end{table*}

We compute the adsorption (or binding) energy, $\Delta E_{ads}$, for each 
molecule/gold composite as defined in  \ref{eq:binding}.
We also computed the elastic deformation energy of the adsorbed molecules, defined as
$\Delta E_{d} = E_{free}-E_{adsorbed}$; $E_{free}$ is the total energy of the free,
relaxed molecule, while $E_{adsorbed}$ is the total energy of the molecule but deformed by the 
adsorption on the surface.
We found that the values of $\Delta E_{d}$ are less than 0.01 eV for all configurations investigated, which is close 
to the numerical accuracy. In a first approximation we can consider that the molecules are not deformed by the adsorption
on the surface - which is in agreement with the weak physisorption model.

\begin{table*}
\caption{ Adsorption energy ($\Delta E$) with (left) and without (right) BSSE correction included,
 for benzene, pyridine and  thiophene on Au(111) and integral over $\delta \rho $
(with and without BSSE corrections),
for benzene, pyridine and thiophene on Au(111).
From top to bottom, the values for the three functionals are reported.}

\label{tbl:5}
\begin{tabular}{l|rrrrr}
\hline
                      & B/Au  & P$_\parallel$/Au& P$_\perp$/Au & T$_\parallel$/Au &T$_\perp$/Au \\
\hline
\hline
 $\Delta E^{LMKLL}$ [eV]  & -0.39/-0.55 & -0.35/-0.52 & -0.30/-0.47 & -0.38/-0.53 & -0.15/-0.25  \\
 $\Delta E^{DRSLL}$ [eV]  & -0.29/-0.68 & -0.28/-0.64 & -0.28/-0.64 & -0.28/-0.64 & -0.15/-0.54  \\
 $\Delta E^{KBM}$ [eV]  & -0.33/-0.49 & -0.31/-0.44 & -0.30/-0.45 & -0.33/-0.44 & -0.16/-0.23  \\
\hline
 $n_e ^{LMKLL}$              &  -0.05/-0.02 & -0.03/-0.01 & -0.06/-0.06 & -0.06/-0.03 & -0.02/-0.01 \\
 $n_e ^{DRSLL}$              &  -0.03/0.01  & -0.02/-0.00 & -0.07/-0.06 & -0.03/-0.02 & -0.02/-0.01 \\
 $n_e ^{KBM}$                &  -0.04/-0.02 & -0.02/0.00 & -0.07/-0.06 & -0.04/-0.02 & -0.02/-0.01 \\
\hline

\end{tabular}
\end{table*}

Let us remind that the experimental values for binding energy of 
benzene on Au(111) is -0.63 eV \cite{benzen_exp}, while for thiophene
a nearly flat configuration with a binding energies between -0.57 eV and -0.68 eV where reported \cite{Liu1, Nambu}.
By inspecting the values in Table 5 we notice the  qualitative agreement between the computed
and experimental results.
Nevertheless the results that do not include  BSSE correction are in good agreement to experimental values.
We recall in this context that it was already pointed out that the counterpoise correction to
BSSE tend to overcorrect the binding energy in the case of dispersion interaction \cite{Sherrill}. 
Indeed, the results obtained by using plane wave calculations \cite{Wei1} are in better agreement with the
experimental values, suggesting that the counterpoise correction is over-correcting the binding energies
in Table 5. Also, as already mentioned, the absence of a dipole moment of the surface, or the
small inaccuracies in the estimation of the charge transfer (see below) can be responsible for the discrepancies
between the plane-waves and PAO calculations.

We note that the binding energy of thiophene in the flat-lying configuration
is about 0.1 eV lower that the binding energy of the perpendicular orientation. This small difference
between the two binding energies corroborates with the experimental fact that at small coverage rate
the flat-lying orientation is preferred while the increase of coverage rate leads to a perpendicular orientation
of the molecules (i.e. a small energy barrier exists between these two configurations).
The same qualitative conclusion appears for pyridine, where an energetic stabilization of 0.2 eV occurs for
the flat-lying orientation.

Overall, LMKLL predict the strongest molecule-surface binding for all configurations, while the DRSLL
leads to the smallest values. Nevertheless, the largest differences are less than 0.1 eV.

\subsection{Molecule-surface charge transfer}

Let us now discuss the charge redistribution in the adsorbates as a function of the exchange correlation functional. 
We analyze the spatially resolved  charge-density difference between the relaxed molecule-surface composite system and 
the sum of its parts,  $\Delta \rho (\vec r)$.
To  this end we compute the quantity:
\begin{equation}
\Delta \rho (\vec r) = \rho_{ads/subs}(\vec r)-[\rho_{ads}(\vec r)+\rho_{subs}(\vec r)]
\label{d_rho}
\end{equation}
where $\rho_{ads/subs} (\vec r)$, $\rho_{ads}(\vec r)$ and $\rho_{subs}(\vec r)$ are the (negative) charge
densities of the relaxed adsorbate-substrate system, adsorbate without substrate and clean surface,
respectively.

In order to compute $\rho_{ads}(\vec r)$ and $\rho_{subs}(\vec r)$ quantities we used two methods. First, we 
remove a part of the system (i.e. the metal or the organic atoms) and compute the densities for the remaining atoms.
Second, we keep the whole system as resulting after the relaxation, but we set as ghost atoms the component atoms for each of the two parts 
(i.e. gold substrate and molecule). The idea is similar to that used in the
Boys-Bernardi counterpoise method \cite{BB}. Indeed, the quantity $\Delta \rho (\vec r)$ is expected to be affected too
by the superposition of the basis sets of the gold and molecule leading to an error in the charge density
in the same way as the binding energy is affected. For this reason we call the resulting values the BSSE corrected values,
although this is an abuse of language.

First, we estimate the molecule-surface charge transfer by
integrating $\Delta \rho(\vec r)$ over selected spatial domains.
Precisely, the ad-structure is divided into two regions by a plane parallel to the metal's surface.
The former one contains the ad-molecule and the latter one contains the surface.
All necessary information are extracted from the grid representation of charge density in real space
as provided by SIESTA \cite{Siesta}. The charge corresponding to the grid point $i$ is computed as
$Q_i = \Delta \rho_i (\vec r) \delta V$, where $\Delta \rho_i (\vec r)$ is the value of $\Delta \rho(\vec r)$
stored at the point $i$ on the grid, and $\delta V$ is the volume associated with a point on the grid.
The numerical results obviously depend on the way the adsorbate system is divided into the two regions
(the position of the separating plane). Nevertheless, the method allows to give a quantitative estimation of
the importance of the charge transfer interaction occurring between molecule and surface.
We define the two regions (molecule and surface) by choosing a plane parallel to the metal's surface located
close to the plane that bisects the ad-system at the vertical position.
The distances from Au(111) surface to separation planes range between 1.90 \AA\ (in the case of pyridine oriented in a 
flat-lying manner for DRSLL functional) and 1.35 \AA\ (the same molecule,
oriented perpendicular to the surface for KBM functional).  

The values for the integral $\Delta \rho (\vec r)$ over the molecule region are given in Table 5;
it can be seen that the results are all negative and close to zero for all models.
Nevertheless, we already pointed out the fact that the molecule-surface distance is overestimated.
For this reason we think that the results in Table 5 are probably underestimated. We computed the value of charge transfer
by using the Model 1 presented above (with KBM exchange-correlation functional) since in this case, the molecule-surface
distance was 2.9 \AA\ (i.e. very close to the experimental values). The resulting value for the number of electrons
in the molecular region was close to 0.2. While this result can be criticized (see the comments in the previous section)
it suggests that the molecule-surface charge transfer is indeed underestimated. Consequently, the 
electrostatic interaction between molecule and surface is expected to be also underestimated.

If the ghost atoms are used to compute the charge densities in equation \ref{d_rho} there is a clear trend to obtain values 
about 100\% larger compared to the initial ones. We think that  a better description of the geometric properties of the adsorbate 
(i.e. the molecule-surface distance) may lead
to a better estimation of the molecule-surface charge transfer. Furthermore, this is expected to change the physics of the
long range interactions and ultimately, the description of the adsorption process for aromatic molecules.

\begin{figure}
\includegraphics[width=8cm, angle=0]{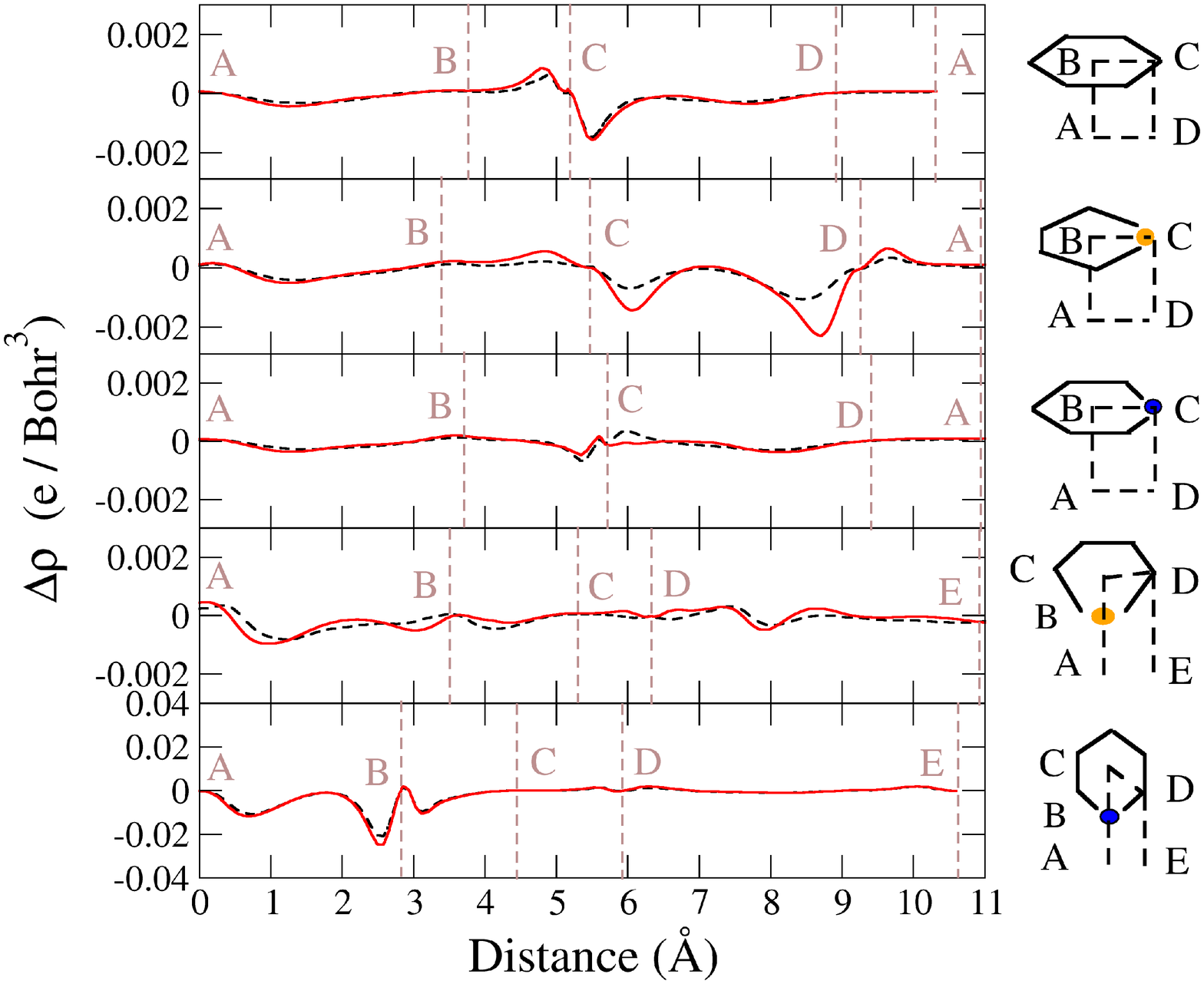}
\caption{Left: values of $\Delta \rho (\vec r)$ computed along selected paths for the molecules adsorbed
on Au(111) surface. Red lines: BSSE corrections included. Black-dotted lines: without BSSE corrections.
 Right column: schematic representation of the paths used in the analysis.
All paths start from the gold surface (point A). We used a blue/dark yellow
circle to symbolize the carbon/sulphur atoms in pyridine and thiophene. }
\label{fig:rho111}
\end{figure}

Next, we go a detailed investigation of  $\Delta \rho (\vec r)$  for selected regions of the space. To this end, we define a close 
path and we plot the values of $\Delta \rho (\vec r)$ along it (i.e. similar to the plots used to represent
the band structure in solids). 
The results for LMKLL functional together with the schematic representation of the paths are represented in \ref{fig:rho111}.
(For the other two exchange correlation functionals we get very similar plots - pictures not shown here).

For flat-lying molecules it can be seen that along the AB segment (i.e. from surface to the center of mass of the molecules) all systems
display negligible variations of the charge density. 
On the plane of $\pi$ rings (BC segment) small fluctuation of the charge occurs in all cases. While for benzene and thiophene
we see an increase of the charge, the pyridine display a decrease of charge density in the vicinity of the nitrogen atom.
The CD segment (i.e. between an atom of the $\pi$ ring and the surface) displays the most important difference between the three curves. 
This is no surprise, since the C point is occupied by different chemical species (carbon for benzene, sulphur for thiophene and
nitrogen for pyridine). For benzene and thiophene a relatively important (i.e. 0.002 e/Bohr$^3$) decrease of the charge density occurs just below
the carbon/sulphur atoms. In the case of pyridine, an increase is present but the values are rather small. For thiophene
a second important fluctuation of $\Delta \rho (\vec r)$ occurs at the gold surface (about 0.003 e/Bohr$^3$ decrease). This is probably caused
by the small chemical interaction occurring between gold and sulphur. In the plane of the gold surface (DA segment) thiophene is the only 
molecule that induces some fluctuations of the electric charge (an increase of about 0.001 e /Bohr$^3$) while for benzene and pyridine the
curves are almost flat in this region, indicating the absence of all chemical interactions.

By comparing the BSSE corrected and uncorrected values we notice that the shape of the two curves are quite similar. 
Slightly different values occurs for BSSE corrected calculations (i.e. few percents smaller). 
For thiophene, relatively large variations occur, indicating that for this system the basis set superposition has the most important effect on
$\Delta \rho (\vec r)$. We also note that the differences present in \ref{fig:rho111} does not account 
for the factor of two obtained previously by integrating the total $\Delta \rho (\vec r)$ in the molecular region. 
This indicates that in addition to the 
differences in the values presented in figure \ref{fig:rho111}, the spatial distribution of density is influenced by 
the BSSE effects on the $\Delta \rho (\vec r)$.

For molecules oriented perpendicular to the surface we found the most important fluctuations in the charge density
along the surface-molecule (AB) segment.
Moreover, for pyridine the values are one order of magnitude larger than the values typically obtained in all other cases.
Precisely, just below the nitrogen atom we found a maximum value of -0.03 e/Bohr$^3$, while close to the surface
the value is about -0.015 e/Bohr$^3$. For comparison, the curve for thiophene has a similar shape, but the value
of $\Delta \rho (\vec r)$ at the surface reaches only -0.001 e/Bohr$^3$.
Next, we see relative flat curves for CD and DE paths in the case of pyridine. For thiophene, we see a 
charge fluctuation on the DE path (i.e. molecule-surface), but the values are still negligible.

In this case too, the BSSE correction to $\Delta \rho (\vec r)$ does not lead to any qualitative changes in the shape of
final curves. This is consistent with our previous conclusion that the main effect of basis set superposition
on the spatial density is to produce variations of the electronic density over wide regions in space, rather than
producing important changes on the localized regions.



\section{Conclusion}

In this paper we numerically addressed specifics of adsorption of benzene, thiophene and pyridine on Au(111) surface. 
Our calculations are based on DFT and the newly developed exchange-correlation functionals including the effect of van der Waals interactions by using localized basis sets and norm-conserving potentials.

Our results indicate that the three tested functionals have the tendency to overestimate the geometric parameters for both gold bulk and surface as well as the molecule-surface distance. 
While the use of localized basis sets based on short-ranged pseudo-atomic orbitals (PAOs) can partially overcome this problem, we found that such computational strategy fails in describing correctly the binding energies of the adsorbates. The main reason for this failure is the existence of an important BSSE effect. On the other hand, 
molecule-surface distance is in good agreement with the experimental data as long as short-ranged PAOs are used. 
By using  geometric models with gold atoms pinned to their bulk position and long range PAOs for gold we obtain a significant decrease of the BSSE effect, as well as a good convergence of the results with respect to basis set size.
The resulting binding energies reach up to 70\% of the experimental value; the best agreement is obtained for LMKLL functional. Nevertheless, the differences between the binding energies computed with the three functionals are very small (less than 0.1 eV).
Furthermore, the model can account for the flat lying orientation of the pyridine and thiophene at low coverages, predicting a differences of 0.1 - 0.2 eV between the corresponding binding energies.
However, the relative position of the molecule with respect to the surface is overestimated with more than 0.6 \AA\ in this case.
Consequently, the molecule-surface charge transfer is expected to be underestimated (i.e., we believe that the obtained value of approximately 0.05 e is most likely unrealistic) and to further affect the molecule-surface interaction energy.

While, in principle, DFT provides an exact description of the electronic ground states, in practice the exchange-correlation functionals and basis sets used in calculations limit its predictive power. 
The present study on model systems clearly show that by using  van der Waals
exchange-correlation functionals and localized basis set  to describe the adsorption of small aromatic molecules 
on Au(111) a qualitative agreement with the experiment can be reached. However, the quality of the results is 
limited by the interplay between the BSSE effect on one hand, and the overestimation of the distances for both bulk 
and molecule-surface relative position on the other hand. These last findings imposed the study of several
geometric models for the molecule-metal systems, in order to overcome the limitations of the functionals in the
study of metallic bulk and surface. Our results describe the effect of all these approximations over the final results
and suggest that an improvement of geometric predictions produced by the van der Waals functionals is expected 
to reflect directly in the physics  of the  small aromatic molecules adsorbed on gold.
 
\section*{Acknowledgements}
 
All the calculations were performed at the Data Center of INCDTIM, Cluj-Napoca.


\begin{thebibliography}{99}
\bibitem{model3} We note here that we discarded the study of a model similar to Model 3 but for an experimental bulk parameter for gold since our preliminary studies indicate a non-realistic behavior of the gold atoms in the surface. Precisely, an important corrugation of the surface due to the squeezing on lateral directions is present while the molecule-surface interaction was found to be less than 0.1 eV. Such a situation is clearly unrealistic, therefore we did not performed further investigation this model.
\end{thebibliography}
\end{document}